\input epsf.tex
\documentstyle[12pt]{article}

\def\.{\!\cdot\!}
\def\:{\!\cdots\!}
\def\[{\left[}
\def\]{\right]}
\def\({\left(}
\def\){\right)}

\def\bk#1{\langle#1\rangle}

\def\h{{1\over 2}}

\def\ket#1{|#1\rangle}

\def\p{\partial}
\def\r2{\sqrt{2}}

\def\be{\begin{eqnarray}}
\def\ee{\end{eqnarray}}

\begin{document}

\rightline{McGill/97-03}
\rightline{UK/97-04}
\vspace{.5 cm}
\begin{center}
{\Large\bf Consistency of the Baryon-Multimeson Amplitudes for Large-$N_c$
QCD}\\
\vspace{.5 cm}
{C.S. Lam$^*$ and K.F. Liu$^\dagger$}\\
\bigskip
{\it $^*$Department of Physics, McGill University,}\\
{\it 3600 University St., Montreal, Q.C., Canada H3A 2T8}\\
{\it Email: Lam@physics.mcgill.ca}\\
\smallskip
{\it and}\\
\smallskip
{\it $^\dagger$Department of Physics and Astronomy,
University of Kentucky}\\
{\it Lexington, KY 40506, U.S.A.}\\
{\it Email: Liu@ukcc.uky.edu}
\end{center}

\begin{abstract}
We study the pion-baryon scattering process $\pi+B\to (n-1)\pi+B$ 
in a QCD theory with a large number ($N_c$) of 
colors.
It is known that this scattering amplitudes decreases
 with $N_c$ like $N_c^{1-n/2}$, 
and that its individual 
tree diagrams grow like $N_c^{n/2}$. The only way these two can
be consistent is for
$n-1$ powers of $N_c$ to be
cancelled when the Feynman diagrams are summed. 
We prove this to be true in tree order for any $n$.
\end{abstract}

\section{Introduction}
QCD with a large
number ($N_c$) of colors \cite{1,2,3} is a beautiful theory, more so
because its mesons and baryons bear an uncanny resemblance to
the real hadrons inspite of such a drastic
assumption on the number of colors present. 
The only cloudy issue had been related to the consistency of the coupling
between mesons and baryons, an issue which has been considerably 
clarified in recent years, thus lending credence to large-$N_c$
phenomenological applications \cite{4,5,6}.

At first sight
the physical attributes of these large-$N_c$ baryons \cite{2,3} 
look very different from the
real ones, as well as the large-$N_c$ mesons. They contain $N_c$ quarks,
whose various spin and isospin alignments produce a large number of
baryon resonances, all with masses proportional to $N_c$.
Since the emission of a pion may flip the spin and isospin of a quark,
these resonances are coupled together into a multi-channel problem.
Moreover, it is known that
the $n$-meson amplitude is proportional to $N_c^{1-n/2}$, both in the
zero-baryon and the one-baryon sectors \cite{2,3}.
Thus all couplings between mesons are weak, decreasing as some powers
of $1/\sqrt{N_c}$, but the Yukawa coupling of a pion to a baryon is
strong and proportional to $\sqrt{N_c}$. This again marks the
difference between large-$N_c$ mesons and baryons.

The strength of this Yukawa coupling produces a number of 
serious problems. It implies that an $n$-meson tree diagram 
in the one-baryon sector
is proportional to $N_c^{n/2}$, because this diagram
 contains $n$ Yukawa coupling
constants and because all baryon propagators are of $O(1)$. 
Not only does it generate undesirably large loop corrections,
it utterly disagrees with the rule that
an $n$-meson amplitude should decrease with $N_c$ like $N_c^{1-n/2}$,
for any $n$ and for any number of loops \cite{2,3}. 
Unless $n-1$  powers of $N_c$ 
are cancelled in the sum of the $n!$ tree diagrams,
the large-$N_c$ rule in the one-baryon sector will not
be self-consistent even in the tree approximation. 

By demanding these cancellations to take place for $n=2$ and $n=3$,
one obtains a set of conditions whose solution
leads to interesting relations satisfied by physical baryons \cite{4}.
These are constraints consistent with quark-model results, but now
obtained without the explicit assumptions of the model. 
In particular, it demands
the presence of a tower of baryon resonances with equal
spin $J$ and isospin $I$, ranging in values form $\h$
to $\h N_c$ (assuming $N_c$ to be odd).  It has
a rotational mass spectrum with
a moment of intertia proportional to $N_c$.
The presence of the baryon resonances are instrumental in effecting
the cancellations needed for the consistency. Similar results were
also obtained from the strong-coupling theory \cite{7} and the 
Skyrme Model \cite{8}.

Alternatively, one can derive the same physical relations from
an explicit quark picture at large $N_c$ \cite{4},
but then one must demonstrate these cancellations
to take place for the sake of consistency.
This has  been carried out in the literature
for $n=2$ and $n=3$ by direct calculations \cite{4}.

These cancellations are progressively more difficult
to achieve for larger $n$ because $n-1$ powers of $N_c$ must be
cancelled.
To complicate the matter further, vertices for pion
emissions are matrices coupling together all the baryon resonances. 
For these reasons it is not
very hopeful to be able to demonstrate the cancellation for large
$n$ by straight
forward computation in the usual way. However, by using a resummation
technique recently developed \cite{9,10}, such cancellations
can be established for {\it tree diagrams} very easily, and
it is the purpose
of this article to discuss how this is done.

The cancellation mechanism leading to the consistency is actually
a rather general phenomenon, not confined to large-$N_c$ QCD. It stems
from a destructive interference of the multi-meson amplitude, valid
even when the mesons are offshell. It is this destructive interference
that suppresses high powers of $N_c$, and it
is the same destructive interference in 
high energy elastic scattering of quarks
that suppresses high powers of  $\ln s$ 
to enable the eikonal and the Regge pictures to be applied, 
and unitarity to be restored. \cite{10,11}.

It should be noted however that
the argument presented here in Sec.~3 is not sufficient to account
for all the necessary cancellations in
loop diagrams, so the consistency for loop amplitudes remains
an open and challenging question. The cancellation for tree
amplitudes may be viewed as a destructive interference between the
external pions, brought on by their Bose-Einstein statistics.
In loop amplitudes internal pions also participate in the inteference
but even so sufficient amount of cancellation will not be attained.
From the known examples of one-loop \cite{3,4} and two-loop
\cite{12} cancellations in the one-pion sector, one sees that counter terms
coming from renormalization are also necessary to effect the desirable
cancellations. This somewhat complicates the physics and alters the
combinatoric nature of the problem, which is why we are yet unable
to extend our result to loop diagrams.

\def\D{\Delta}
\def\s{\sigma}
\def\ss{[\s_1\s_2\cdots\s_n]}

\section{Nonabelian Cut Diagrams}
The resummation mentioned above replaces the sum
of Feynman tree diagrams (Fig.~1(a)) with the sum of
{\it nonabelian cut diagrams} (Fig.~1(b)) \cite{9, 10}. The latter
are organized in such a way that the interferences of 
Bose-Einstein amplitudes are automatically built in.
With that tool the proof of the consistency criterion follows 
almost immediately.

The resummation theorem applies to any tree amplitude whose main trunk
carries a large energy, either in the form of a large mass
as in the present case of large $N_c$, or a large
kinetic energy as in the case of high-energy quark-quark elastic
scattering. The energies and momenta of the emitted bosons are 
comparatively small, but they can be offshell to allow this tree amplitude
to be sewed up to others to form a loop diagram. In this way
the resummation theorem and the resulting nonabelian cut diagrams 
are applicable in the presence of loops as well.
 
Tree diagrams will be labelled by the order
their meson lines appear along the baryon trunk. The tree diagram in Fig.~1(a)
for example will
be denoted by $[231465]$.
We will construct  from each
Feynman tree diagram a {\it nonabelian cut diagram} \cite{10}
by placing cuts on
some of its propagators as follows. A cut is put
after a meson line iff there is no meson to its right
designated by a smaller number. Denoting
a cut diagram by a subscript $c$, and indicating a cut
by a vertical bar, the cut diagram for Fig.~1(a) is
$[231465]_c=[231|4|65]$, as shown in Fig.~1(b).

\begin{figure}
\vskip -8 cm
\centerline{\epsfxsize 4.7 truein \epsfbox {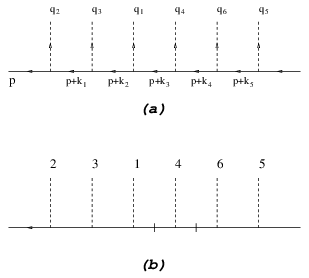}}
\nobreak
\vskip -1.5 cm\nobreak
\caption{(a) A Feynman
tree diagram for baryon-meson scattering; (b) the nonabelian
cut diagram corresponding to (a).}
\end{figure}

Let $p$ be the final momentum of the baryon and $q_i$ be the outgoing
meson momenta. Using $k_i=\sum_{j=1}^iq_{\s_j}$
to denote the sums of meson momenta for the diagram $[\s_1\s_2\cdots \s_n]$,
the momentum of the $i$th baryon is then $p+k_i$.
The assumption of the tree trunk carrying large energy is used to
approximate the $i$th baryon propagator
\be {M_i+\gamma(p+k_i)\over (p+k_i)^2-M_i^2+i\epsilon}
\simeq
\h(1+\gamma^0){1\over k_i^0-\D M_i+i\epsilon}, \ee
 where $\D M_i
=M_i-M$ is the mass difference between the baryon resonance and
the nucleon.
Implicit in this is the assumption that while $M$ and $M_i$
are $O(N_c)$, the difference $\D M_i$ is at most
$O(1)$ as $N_c\to\infty$ in order to keep all the baryons
at a constant velocity.

With this approximation, the Feynman amplitude for
$\ss$ is given by
\be A\ss=\h(1+\gamma^0)a\ss V\ss,\ee
where $V\ss=V_{\s_1}V_{\s_2}\cdots
V_{\s_n}$ is simply the product of all the vertices $V_i$, and
we have assumed that the projection operator $\h(1+\gamma^0)$
commutes with the vertex operators $V_i$ attached to the
$i$th meson line. For the moment we will also assume that
all $\D M_i=0$, but this restriction will be lifted later.
The spacetime part of the amplitude in (2) is then given by
$a\ss=-2\pi i\delta(\sum_{i=1}^n q_i^0)
\prod_{i=1}^{n-1}(k_i^0+i\epsilon)^{-1}$.

On the other hand, 
the amplitude for a {\it nonabelian cut diagram} is defined as follows.
\be A\ss_c=\h(1+\gamma^0)a\ss_c V\ss_c,\ee
where $a\ss_c$ is obtained from $a\ss$ by replacing the
Feynman propagator $(k_i^0+i\epsilon)^{-1}$ of a cut line
by the Cutkosky cut propagator $-2\pi i\delta(k_i^0)$.
The vertex part $V\ss_c$ is obtained from $V\ss$
by replacing the product of $V_i$'s straddling 
{\it uncut lines} by (multiple) 
commutators. For example,
$V[231465]_c=V[231|4|65]=[V_2,[V_3,V_1]]V_4[V_6,V_5]$.

The resummation formula (called the {\it multiple commutator formula} in \cite{9}) asserts that the sum of
the Feynman amplitudes is equal to the sum of the nonabelian cut
amplitudes,
\be  \sum_\s^{n!}A\ss=\sum_\s^{n!}A\ss_c, \ee
when the sum is taken over all the $n!$ permutations
$\s=[\s_1\s_2\cdots\s_n]$ of $[12\cdots n]$.

In the special case when the vertices $V_i$ are abelian so they mutually
commute, the only surviving term is the one without any commutator
appearing, which is given by the cut diagram with every baryon 
propagator cut. The spacetime part $a[123\cdots n]_c$ is now a product
of $\delta$-fucntions in $q_i^0$, showing neatly a very peaked interference
pattern in all the variables $q_i^0$. 
Away from $q_i^0=0$, the interference is purely destructive.

In case of nonabelian vertices,
the different terms on the right-hand side of (4)
carry different internal quantum numbers, and
their spacetime parts exhibit varying degrees of destructive
interference according to the number of $\delta$-functions present.
However, since the number of $\delta$-functions plus the number
of commutators is the same for every term, what is lacking in spacetime
destructive interference is made up by 
the `destructive interference' in the internal quantum
numbers, in the following sense. Imagine $V_i$ to be the generators
of a Lie group in a low-dimensional representation. Then products of
$V_i$ will contain progressively higher-dimensional representations and
hence larger quantum numbers, but commutators of them will simply behave
like a single $V_i$, creating only small quantum numbers. In this sense
commutators represents an `interference' in which large quantum numbers
tend to be wiped out. We shall see that it is this kind of `interference'
that suppresses the high powers of $N_c$.

These formulas, suitably modified,  are applicable even when baryon mass
degeneracy is lifted, provided we insert into the tree
new vertices $V_i'=\Delta M$ carrying away no
energy. To see that, let
$\D M$ be
the diagonal operator whose matrix elements are the mass
differences $\D M_i$. Using the expansion
\be{1\over k_i^0-\D M+i\epsilon}={1\over k_i^0+i\epsilon}\sum_{m=0}^\infty
\({\D M\over k_i^0+i\epsilon}\)^m, \ee
we see that the new vertex necessary is simply $V_i'=\D M$.

\section{Proof of the Consistency Criterion}
Let us first review some standard facts in the one-baryon sector \cite{5}.
The quark-pion interaction is proportional to
$N_c^{-\h}\bar\psi\gamma^\mu\gamma_5\tau^a\p_\mu\pi_a$, in which
the coefficient $N_c^{-\h}$ is fixed (see later) by the requirement
of the meson-baryon Yukawa constant being of order $\sqrt{N_c}$.
In the rest frame of the baryon, the large component $\phi$ of the
Dirac spinor $\psi$ dominates, so this interaction is
reduced to an expression proportional
to $N_c^{-\h}\{\sigma^i\tau^a\}\p_i\pi_a$, where $\{\Gamma\}\equiv
\phi^\dagger\Gamma\phi$. This
in turn determines the pion-baryon vertex to be proportional to
$V_i=N_c^{-\h}\{\s^i\tau^a\}$, with `$a$' labelling the isospin of the
pion it couples to.

The large-$N_c$ rules for $n$-meson amplitude in the one-baryon sector were
derived from the quark picture using the Hartree approximation \cite{2}.
In this approximation, the wave function of a baryon state 
$\ket{B_{J,I}}$ with spin $J$ and
isospin $I$ can be represented by an
$SU(2)_J$ and an  $SU(2)_I$ Young tableau, as shown in Figs.~2(a)
and 2(b). For this color-singlet and $s$-state baryon
to have a totally antisymmetric wave function, the spin and isospin
tableax must be identical, which implies $I=J$.
The double boxes appearing in
a column of the tableau
are singlets of quark pairs in $J$ or $I$, so they are killed by the spin
operator $\{\sigma^i\}$ and the isospin operator $\{\tau^a\}$.
However,
$\{\s^i\tau^a\}\not=\{\s^i\}\{\tau^a\}$, so these singlets are not
killed by $\{\s^i\tau^a\}$. Since there are $O(N_c)$
columns in a tableau,
the baryon matrix element $\bk{V_i}$ is of order
 $N_c^{-\h}\.N_c=\sqrt{N_c}$, as it should for a Yukawa coupling constant.
For simplicity, the notation $\bk{{\cal O}}\equiv \bk{B_{J',I'}|{\cal
O}|B_{J,I}}$ has been used.

\begin{figure}
\vskip -4 cm
\hskip3cm\centerline{\epsfxsize 4.7 truein \epsfbox {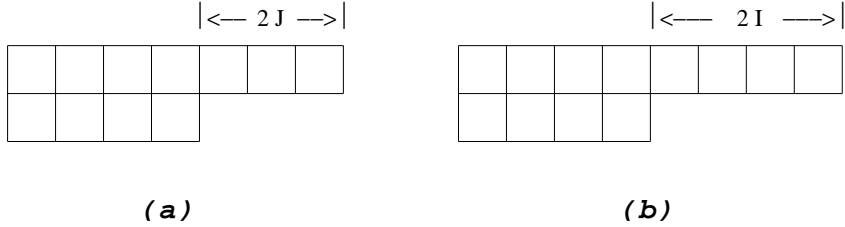}}
\nobreak
\vskip -6 cm\nobreak
\vskip -2.5 cm
\caption{Young tableaux for (a) $SU(2)_J$ and (b) $SU(2)_I$.}
\end{figure}

Similarly, the matrix element of a two-body operator, $\bk{\{\s^i\tau^a\}
\{\s^j\tau^b\}}$, is of order $N_c^2$, but their commutator is only
of order $N_c$. This is so because $[\{\Gamma_1\},\{\Gamma_2\}]=
\{[\Gamma_1,\Gamma_2]\}$, and because
\be [\s^\mu\tau^\alpha,\s^\nu\tau^\beta]=\h[\s^\mu,\s^\nu]
[\tau^\alpha,\tau^\beta]
+\h[\s^\mu,\s^\nu]_+[\tau^\alpha,\tau^\beta]_+\ee
is a linear combination of $\s^\lambda\tau^\gamma$, thus making
its matrix elements order $N_c$.
This means that  $\bk{[V_i,V_j]}$ is of order $(N_c^{-\h})^2N_c=1$.
Similarly, each time an additional commutator appears, the matrix 
element is reduced by an additional power of $N_c^{-\h}$. In particular,
the matrix element of an $n$-tuple commutator is of order
$N_c^{1-n/2}$.
In these expressions, 
$\s^0$ and $\tau^0$ are respectively the unit  matrices in the spin
and isospin spaces.

We proceed now to prove the consistency criterion for the pion-baryon
scattering amplitude $\pi+B\to(n-1)\pi+B$. 
We shall first assume all $\D M_i=0$
and all pions to be coupled directly to the baryon. We will also
take the pion mass to be non-zero.

One of the $n$ pions is incoming and the remaining $n-1$ are outgoing,
so one of the $q_i^0$ is negative but the rest of them are positive.
Energy conservation, or the requirement of the initial baryon to be
on-shell, demands that the sum of the $n$ $q_i^0$'s to be zero. However,
because all but one of them are positive, a partial sum of them can
never be zero, which is to say that the only surviving terms in (4)
are the ones without any Cutkosky cut. These incidentally are 
the nonabelian cut diagrams
with pion 1 at the far right. For such terms,
the vertex factor $V\ss$  contains $n$-tuple commutators
of the vertices $V_i$, 
whose baryon matrix elements are of order $N_c^{1-n/2}$ as 
we saw before.
This then shows that the sum of the $n!$ tree diagrams is of
order $N_c^{1-n/2}$, and we have attained just the right amount of 
cancellations required by consistency.

This conclusion remains valid
 without the special assumptions. If $\D M_i\not= 0$,
then rotational invariance demands it to be of the form
$\D M=c\{\vec \s\}\.\{\vec \s\}/N_c$ \cite{5}, so
$[\D M, \{\s^i\tau^a\}]=(2c/N_c)\{\s^i\tau^a\}$. 
This means that any commutator with $\D M$ will only lead to
subleading dependences at large $N_c$. The same will also be true
if some of the mesons are coupled directly to other mesons rather than
the baryon, because all meson couplings vanish as a power of $N_c^{-\h}$.
Seagull type of diagrams are also negligible.

Before ending, 
we should also remark on the special situation when the pion is massless. 
In that case pion energies can be zero and the $\delta$-functions
in the partial sums of $q_i^0$ can no longer be thrown away so easily.
These terms have
less commutators of $V_i$ and hence higher powers of $N_c$ than
$N_c^{1-n/2}$. However, these correspond exactly
to the terms in which different pions hit different quarks \cite{2,3},
rather than the same quark which leads to the familiar dependence  
of $N_c^{1-n/2}$.

This work is supported by the Natural Science and Engineering
Research Council of Canada, and the Fonds pour la Formation de
Chercheurs et l'Aide \`a la Recherche of Qu\'ebec (CSL), and
by USDOE grant DE-FG05-84ER40154 (KFL). CSL would like to thank
Markus Luty and Greg Keaton for stimulating discussions, and Y.J. Feng
for drawing the diagrams.


\begin{thebibliography}{9}
\bibitem{1} G. 't Hooft, {\it Nucl. Phys.} {\bf B72} (1974) 461.
\bibitem{2} E. Witten, {\it Nucl. Phys.} {\bf B160} (1979) 57.
\bibitem{3} S. Coleman, Erice Lectures (1979).
\bibitem{4} A.V. Manohar, Plenary Talk at PANIC 96, 
hep-ph/9607484;
R.F. Dashen, E. Jenkins and A.V. Manohar, {\it Phys. Rev. D}
{\bf 49} 4713; 
R. Dashen and A.V. Manohar, {\it Phys. Lett. B} {\bf 315} (1993) 425, 438;
E. Jenkins, {\it Phys. Rev. D} {\bf 55} (1997) 10; 
{\it Phys. Lett. B} {\bf 315} (1993) 431, 441, 447. 
\bibitem{5} C. Carone, H. Georgi, and S. Osofsky, {\it Phys. Lett. B}
{\bf 322} (1994) 227; 
M.A. Luty and J. March-Russell, {\it Nucl. Phys.}
{\bf B426} (1994) 71; M.A. Luty, {\it Phys. Rev. D} {\bf 51} (1995) 2322;
M.A. Luty, J. March-Russell and M. White, {\it Phys. Rev. D} {\bf 51} (1995)
2332; R.F. Dashen, E. Jenkins and A.V. Manohar, {\it Phys. Rev. D}
{\bf 51} (1995) 3697.
\bibitem{6}
P.B. Arnold and M.P. Mattis, {\it Phys. Rev. Lett.} {\bf 65} (1990) 831;
{\it Phys. Rev.} {\bf 51} (1995) 3267; M.P. Mattis and R.R. Silbar,
{\it Phys. Rev. D} {\bf 51} (1995) 3267.

\bibitem{7} J.-L. Gervais and B. Sakita, {\it Phys. Rev. Lett.} {\bf 52}
(1984) 87; {\it Phys. Rev. D} {\bf 30} (1984) 1795.
\bibitem{8} G.S. Adkins, C.R. Nappi, and E. Witten, {\it Nucl. Phys.}
{\bf B228} (1983) 552; J. Bijnens, H. Sonoda, and M.B. Wise,
{\it Phys. Lett. B} {\bf 140} (1984) 421.
\bibitem{9} C.S. Lam and K.F. Liu, {\it Nucl. Phys.} {\bf B483} (1997) 514.
\bibitem{10} Y.J. Feng, O. Hamidi-Ravari, and C.S. Lam,
{\it Phys. Rev. D} {\bf 54} (1996) 3114 (hep-ph/9604429)
\bibitem{11} Y.J. Feng and C.S. Lam, hep-ph/9606351, to appear in 
{\it Phys. Rev. D}.
\bibitem{12} G.L. Keaton, {\it Phys. Lett. B} {\bf 372} (1996) 150.
\end{thebibliography}
\end{document}